\documentclass[twocolumn,doublespace,showkeys,showpacs,superscriptaddress]{IEEEtran}

\usepackage{graphicx,epic,eepic,epsfig,amsmath,latexsym,amssymb,verbatim,color}

\usepackage{verbatim}

\usepackage{theorem}

\newtheorem{definition}{Definition}

\newtheorem{theorem}[definition]{Theorem}

\def\squareforqed{\hbox{\rlap{$\sqcap$}$\sqcup$}}
\def\qed{\ifmmode\squareforqed\else{\unskip\nobreak\hfil
\penalty50\hskip1em\null\nobreak\hfil\squareforqed
\parfillskip=0pt\finalhyphendemerits=0\endgraf}\fi}
\def\endenv{\ifmmode\;\else{\unskip\nobreak\hfil
\penalty50\hskip1em\null\nobreak\hfil\;
\parfillskip=0pt\finalhyphendemerits=0\endgraf}\fi}

\mathchardef\ordinarycolon\mathcode`\:
\mathcode`\:=\string"8000
\def\vcentcolon{\mathrel{\mathop\ordinarycolon}}
\begingroup \catcode`\:=\active
  \lowercase{\endgroup
  \let :\vcentcolon
  }

\newcommand{\nc}{\newcommand}
\nc{\rnc}{\renewcommand} \nc{\beq}{\begin{equation}}
\nc{\eeq}{{\end{equation}}} \nc{\beqa}{\begin{eqnarray}}
\nc{\eeqa}{\end{eqnarray}} \nc{\lbar}[1]{\overline{#1}}
\nc{\bra}[1]{\langle#1|} \nc{\ket}[1]{|#1\rangle}
\nc{\ketbra}[2]{|#1\rangle\!\langle#2|}
\nc{\braket}[2]{\langle#1|#2\rangle} \nc{\proj}[1]{|
#1\rangle\!\langle #1 |} \nc{\avg}[1]{\langle#1\rangle}
\rnc{\max}{\operatorname{max}} \nc{\Rank}{\operatorname{Rank}}
\nc{\smfrac}[2]{\mbox{$\frac{#1}{#2}$}} \nc{\Tr}{\operatorname{Tr}}
\nc{\id}{\operatorname{id}} \nc{\1}{\openone} \nc{\ox}{\otimes}
\nc{\dg}{\dagger} \nc{\dn}{\downarrow} \nc{\cA}{{\cal A}}
\nc{\cB}{{\cal B}} \nc{\cC}{{\cal C}} \nc{\cD}{{\cal D}}
\nc{\cE}{{\cal E}} \nc{\cF}{{\cal F}} \nc{\cG}{{\cal G}}
\nc{\cH}{{\cal H}} \nc{\cI}{{\cal I}} \nc{\cJ}{{\cal J}}
\nc{\cK}{{\cal K}} \nc{\cL}{{\cal L}} \nc{\cM}{{\cal M}}
\nc{\cN}{{\cal N}} \nc{\cO}{{\cal O}} \nc{\cP}{{\cal P}}
\nc{\cQ}{{\cal Q}} \nc{\cR}{{\cal R}} \nc{\cS}{{\cal S}}
\nc{\cT}{{\cal T}} \nc{\cV}{{\cal V}}\nc{\cX}{{\cal X}}
\nc{\cZ}{{\cal Z}} \nc{\supp}{{\operatorname{supp}}}
\nc{\var}{\operatorname{var}} \nc{\rar}{\rightarrow}
\nc{\lrar}{\longrightarrow}
\nc{\polylog}{\operatorname{polylog}}\nc{\poly}{\operatorname{poly}}

\def\a{\alpha}

\def\e{\epsilon}

\nc{\RR}{{{\mathbb R}}}
\nc{\CC}{{{\mathbb C}}}
\nc{\FF}{{{\mathbb F}}}
\nc{\ZZ}{{{\mathbb Z}}}
\nc{\PP}{{{\mathbb P}}}
\nc{\QQ}{{{\mathbb Q}}}
\nc{\UU}{{{\mathbb U}}}
\nc{\EE}{{{\mathbb E}}}

\nc{\be}{\begin{equation}}
\nc{\ee}{{\end{equation}}}
\nc{\bea}{\begin{eqnarray}}
\nc{\eea}{\end{eqnarray}}
\nc{\<}{\langle}
\rnc{\>}{\rangle}
\nc{\Hom}[2]{\mbox{Hom}(\CC^{#1},\CC^{#2})}
\nc{\rU}{\mbox{U}}

\nc{\ob}[1]{#1}

\def\mynp{\lceil np \rceil}
\def\N{\cN^{\rm adv}_{p,n}}

\begin{document}

\title{Communicating over adversarial quantum channels using quantum
list codes}

\author{
Debbie Leung and Graeme Smith 
\thanks{
Debbie Leung and Graeme Smith were both with the Institute for Quantum
Information, California Institute of Technology, MSC 107-81, Pasadena,
CA 91125, USA.  Debbie Leung is now at the Institute for Quantum
Computing, University of Waterloo, Ontario, Canada N2L 3G1, and Graeme
Smith is now at the Department of Computer Science, University of
Bristol, Bristol, BS8 1UB, UK. Emails:wcleung\@@iqc.ca and
gsbsmith\@@gmail.com . }} 

\maketitle

\markboth{Submitted to IEEE Transactions On Information Theory}
{Capacities}

%
%

\date{\today}


\begin{abstract}
We study quantum communication in the presence of adversarial noise.
In this setting, communicating with perfect fidelity requires a
quantum code of bounded minimum distance, for which the best known
rates are given by the quantum Gilbert-Varshamov (QGV) bound.  Asking only 
for arbitrarily high fidelity and letting the sender and
reciever use a secret key of length logarithmic in the number of
qubits sent, we find a dramatic improvement over the QGV rates.  In
fact, our protocols allow high fidelity transmission at
noise levels for which perfect fidelity is impossible.  To achieve
such rates, we introduce fully quantum list codes, which
may be of independent interest.
\end{abstract}



\parskip=1ex
\parindent=0ex

\section{Introduction}

\label{sec:introduction}
Effectively dealing with noise is a major challenge faced by all
proposals for the coherent manipulation of quantum information.  
Besides communication, sending a quantum state over 
a noisy channel models noisy storage, and as such, characterizing 
communication rates for quantum channels is a 
central question in the study of both quantum information and computation.

Various asymptotic capacities of quantum channels have been studied
\cite{D03,DHW03,DS03,Shor02,W99,W04POVM,BarnumKN00,SW97,Holevo98,Lloyd97,BSST02}. However, this
work has been almost exclusively concerned
with discrete memoryless channels (DMCs), wherein a sender and
receiver use many independent and identical copies of a channel.  In
this scenario, one studies the asymptotic communication rate possible
using an operation of the form $\cN^{\ox n}$, where $\cN$ is the
channel under consideration and the rate is $R = k/n$ where
$k$ is the number of high fidelity logical qubits sent.  Relatively
little is known outside of the DMC scenario, with notable exceptions
found in \cite{BDM03,KW05,HN02,RenThesis,BD06}.

In this paper, we study an adversarial
quantum channel (AQC), which is perhaps as different from a DMC as one
can imagine. When sending $n$ qubits over an AQC, instead of
errors on different qubits occuring independently, an adversary
who knows what protocol is being used tries to foil the
communication by maliciously choosing a superposition of errors,
subject only to a restriction on the number of qubits each error
affects.  We call this channel $\N$, where $p$ is the fraction of
qubits the adversary is allowed to corrupt. $\N$ is the
natural quantum generalization of the classical adversarial channel
that was considered in \cite{Lang04,Guru03} and whose roots go
back to \cite{Hamming50}.

If the receiver must reconstruct the logical state exactly,
communicating over $\N$ requires a quantum error correcting code
(QECC) of distance $2\mynp+1$.  
The quantum Gilbert-Varshamov bound guarantees the existence of such 
a code with a rate of at least $1{-}H(2p){-}2p\log 3$
\cite{Gottesman97}, where logarithms are taken base 2 here and
throughout.  Communication beyond this rate is
possible only if QECCs beating the Gilbert-Varshamov bound exist,
which is a question that has been quite difficult to resolve.
Furthermore, Rains has shown \cite{Rains99,Rains03} that no quantum code
can have distance greater than $n\left( 3-\sqrt{3}\right)/4 \approx 0.317n$, so that it is
impossible to send even a single qubit for $p \geq (3-\sqrt{3})/8 \approx 0.158$.

However,  if we ask only for a high fidelity reconstruction, and
allow the sender and receiver to share a secret key of size $O(\log
n)$ it is possible to communicate at rates much higher than the
Gilbert-Varshamov and Rains bounds suggest.  Below, we present a
coding strategy for this scenario with a rate of $1{-}H(p){-}p\log 3$,
which is significantly larger than the Gilbert-Varshamov rate for all
values of $p$ and remains nonzero up to $p\approx 0.189$.  Our rate
equals the best communication rate via the depolarizing channel (in
the DMCs scenario) of error probability $p$ using only nondegenerate
codes.

There are three ingredients in achieving such rates with negligible
length secret keys.  The first is a predetermined quantum
list code that is known to the adversary.  This alone allows high-rate
but low-fidelity transmission.  To improve the fidelity, a random
subcode is further chosen according to a secret key unknown to the
adversary.  Finally, the subcode is derandomized using small-biased
sets.
%
%

Before explaining the construction of our code, we first discuss some
intuition on why the code works and how resources are being reduced.  
Informally, a quantum list code is an error correcting code with the
relaxed reconstruction requirement that the decoded state be equal to
the original state acted on by a superposition of a small number of
errors. 
We call the number of errors the ``list length.''  
This relaxation allows a considerable increase in rate over QECCs, and
by a random coding argument we show there are list codes with
constant-length lists and rates approaching $1{-}H(p){-}p\log 3$ that
tolerate $pn$ errors.
To distinguish between the errors in the list and communicate with
high fidelity, the sender and receiver select a large subcode of
the list code using a secret key.  In particular, this can be chosen
pseudorandomly by using $O(\log n)$ bits of secret key.

We can interpret our code as a set of (parity) check conditions that
yield syndrome information.  Most of these conditions are used in
list-decoding and can be known to the adversary, and the rest of the
conditions are pseudorandom and with high probability are capable of
completely distinguishing the errors on the list.
Note that there are simpler constructions using randomness unknown to
the adversary, and we now make a comparison.
The first construction is simply a random (nondegenerate) quantum
error correcting code achieving the same rate but requiring $O(n^2)$
bits of secret key \footnote{It is folklore, somewhat implicit in the hashing protocol of
\cite{BDSW96}.  It is also implied by our current construction.}.
Second, one could use a secret permutation of the $n$ qubits in the
AQC turning the adversarial channel to something very similar to $n$
depolarizing DMCs of error $p$ in the DMCs setting
\cite{ShorPreskill00,RenThesis}.
The best known rate is similar to ours (but slightly better for large
$p$) but the cost will be $O(n\log n)$ bits of key, which,
unfortunately, still gives a divergent key rate.
Third, the standard derandomizing technique of key recycling cannot be
used in a straightforward way in the current, adversarial, context.
Thus, our hybrid construction involving a known list-code and a
pseudorandom subcode demonstrates what type of randomness is
unnecessary, and can be seen as a method to derandomize other
key-inefficient protocols, achieving the same task with a much shorter
key.

For the rest of the paper, we summarize related works, review
background material, present the details of our construction, after
which we discuss various applications and open problems.

{\bf Related work} 

Approximate error correction was studied in \cite{Leung97b} to reduce
the block length (and thus improving the rate) for a more specific
error model.  Success criterion and algebraic sufficient conditions
were given.  Reference \cite{Schumacher01} provided an information
theoretic approximate error correction criterion.  The approximation
in these works stems from a {\em relaxed decoding} procedure.  Much
closer to our work is Ref.~\cite{CGS05} (in the context of quantum
secret sharing) that used a randomized code to maximize the distance
with high probability but the rate is low (of lesser concern in that
context).  Our construction was inspired by that in \cite{Lang04} in
the classical setting.  Further comparisons between our work and these
earlier results and insights obtained will be discussed in Section
\ref{sec:disc}.

After the initial presentation of this result \cite{QIP06}, we learned
of two independent studies of list codes, both in settings quite
different from our own.  Ref.~\cite{KY06} studied decoding of {\em
classical list codes} with quantum algorithms, and
Ref.~\cite{Hayashi06} studied list codes for sending {\em classical}
messages via iid quantum channels.

\smallskip

\section{Background and Definitions} 

Our sender, receiver, and adversary will be named Alice, Bob, and Eve,
respectively.  The encoding of a $k$-qubit state $\ket{\psi}$ into a
QECC will be written as $\ket{\bar{\psi}}$.  We call the Pauli group
acting on $n$ qubits $\cG_n$ and write its elements in the form $P =
i^t X^{\mathbf{u}}Z^{\mathbf{v}}$, where $t \in \{0,1,2,3\}$,
$\mathbf{u}, \mathbf{v}$ are binary vectors of length $n$,
$X^{\mathbf{u}}$ ($Z^{\mathbf{v}}$) denotes $X^{u_1}\ox \cdots \ox
X^{u_n}$($Z^{v_1}\ox \cdots\ox Z^{v_n}$), $X = \binom{0 \ 1}{1 \ 0}$
and $Z = \binom{1 \ \ \ 0}{0\ -1}$.  The (anti)commutation relation
between $P_1,P_2 \in \cG_n$ is determined by $P_1P_2 =
(-1)^{\omega(P_1,P_2)} P_2P_1$ with $\omega(P_1,P_2) =
\mathbf{u}_1\cdot\mathbf{v}_2 + \mathbf{u}_2\cdot\mathbf{v}_1$, where
the dot products and sum are computed in arithmetic modulo two.  We
let $\langle P_l\rangle$ denote the subgroup of $\cG_n$ generated by a
set of Pauli elements $\{P_l\}$.

A state $\ket{\psi}$ is said to be stabilized by a Pauli
matrix $P$ when $P\ket{\psi} = \ket{\psi}$. 
An $[n,k]$ stabilizer code is a $2^{k}$-dimensional space of $n$-qubit
states simultaneously stabilized by all elements of a size $2^{n-k}$
abelian subgroup of $\cG_n$.
The abelian subgroup is typically called $S$ and is referred to as the
code's stabilizer, and has $n{-}k$ generators denoted by
$\{S_i\}_{i=1}^{n-k}$. For any $E\in \cG_n$ we refer to the
$(n{-}k)$-bit string $\omega(E,S_i)$ as the syndrome of $E$
\cite{Gottesman97,Nielsen00bk}.  
The weight of a Pauli matrix $P$, which we denote by ${\rm wt}(P)$, is
the number of qubits on which $P$ acts nontrivially, and we call a
stabilizer code an $[n,k,d]$ code if it can detect all errors outside
of $S$ of weight less than the distance $d$, which is equivalent to
being able to correct all errors of weight less than $\lfloor
(d{-}1)/{2} \rfloor$.
%
%
For any positive real number $r$, let $\cE^r$ be the set of Pauli matrices
of weight no more than $\lfloor r \rfloor$.  
Let $N(S)$ be the set of all unitaries leaving $S$ invariant under
conjugation.  ($N(S)$ is the center and also the normalizer
of $S$ in $\cG_n$, thus the symbol $N$.)
Note that two errors $E_i$ and $E_j$ have the same syndrome if and
only if $E_i^\dg E_j \in N(S)$.
Thus $S$ defines an $[n,k,d]$ code exactly when every pair of errors
$E_i,E_j\in \cE^{(d{-}1)/2}$ satisfies $E_i^\dg E_j \not\in N(S)-S$.
Intuitively, it means that the syndrome can be used to identify all
errors of concern up to a multiplicative factor that is in $S$ 
and has no effect on the codespace.

We state a property of the Pauli group that will be useful later.  For
any subgroup $G$ of $\cG_n$, for any set {\sc s} of $i$ {\em
independent} elements in $G$, and a specific (ordered) list of $i$
(anti)commutation relations with elements of {\sc s}, exactly
$|G|/2^{-i}$ elements of $G$ will satisfy those relations.

\begin{definition}
The $n$-qubit adversarial quantum channel with error rate $p$, which
we call $\N$, acts on a state of $n$ qubits, $\rho$, and is of the
form
\begin{equation}\label{Eq:AdvChannelDef}
\N(\rho) = \sum_i A_i \rho A_i^\dg \mbox{ with } A_i =
\sum_{E \in \cE^{pn}} \a_E^i E
\end{equation}
subject to the requirement that $\sum_i A_i^\dg A_i = I$ and where
$\cE^{np} = \{ E {\,\in \,} \cG_n \, | \, wt(E) {\, \leq \,} pn\}$ 
is as defined before.
%
The particular choice of the $\{A_i\}$'s is made by Eve only after
Alice and Bob have decided on a communication strategy.
\end{definition}

Notice that to communicate effectively over $\N$
one must find a strategy that works with high fidelity for
{\em all} channels described by Eq.~(\ref{Eq:AdvChannelDef}).  To do
this, we will use quantum list codes, which are defined below.

\begin{definition}
We say that an $[n,k]$ stabilizer code, ${\cal C}$, is an
$[n,k,t,L]$-list code if there is a decoding operation, $\cD$, such
that for every $E_i \in \cE^{t}$ and $\ket{\bar{\psi}} \in \cC$, the
decoded $k$-qubit state, along with the syndrome $s$, is given
by $\cD(E_i\proj{\bar{\psi}}E_i^\dg) = \sum_s \sum_j A_j^s
\proj{{\psi}} A_j^{s\dg}\ox\proj{s}$ where $\sum_{sj} A_j^{s\dg} A_j^s = I$,
and each $A_j^s$ is a linear combination of the $2^L$ elements of
$\langle P^s_l\rangle_{l=1}^L$, where $\{P_l^s\}_{l=1}^L$ is a list of
logical errors on the codespace and $\langle P^s_l\rangle_{l=1}^L$ is
the group they generate.
\end{definition}

Note that in the above definition, the set $\{P_l^s\}_{l=1}^L$ generating 
the error list depends on the syndrome $s$.  


%

\section{Quantum List Codes} 
We now show that, asymptotically, there exist $[n,k,t,L]$-list codes
with favorable parameters.  We proceed by considering random
stabilizer codes, arguing along the lines of
\cite{BDSW96,Gottesman97}.  In particular, we will show that if we
choose a random stabilizer code with rate as below, in the limit of
large $n$ the probability of it failing to be $L$-list decodable is
less than 1.

\begin{theorem}\label{Thm:ListCodesExist} 
$[n,\lfloor Rn\rfloor,\lfloor pn\rfloor,L]$-list codes exist for
sufficiently large $n$ and for
\begin{equation} 
R < 1 - \left(1 + \frac{1}{L}\right)\left(H(p) + p \log 3\right) \,.
\end{equation} 
\end{theorem}
\begin{proof}
Let $N_E = |\cE^{pn}|$ and $\cE^{pn} =\{E_i\}_{i=1}^{N_E}$.  
Since two errors $E_i$ and $E_j$ have the same syndrome iff
$E_i^\dagger E_j \in N(S)$, a code will fail to be $L$-list decodable
only if there are $L+1$ {\em independent} errors $E_0, \cdots, E_{L}$
outside of $S$ having the same syndrome.  Mathematically, this means
$E_i^\dagger E_j \in N(S)$ for $0 \leq i,j \leq L$ (or equivalently,
$E_0^\dagger E_j \in N(S)$ for $1 \leq j \leq L$).
%
%
The proof consists of two steps: (1) bounding the probability (over
the code) for a fixed list of $L$ independent Pauli matrices to be in
$N(S)$, and (2) taking the union bound over all such possible lists to
show that list-decoding will fail with probability (over the code)
{\em strictly} less than $1$.  Thus, the desired list code must exist.

Step (1) is essentially a counting argument.  How many ways can we
choose $n{-}k$ stabilizer generators $S_1,S_2,\cdots S_{n-k}$?  Here we omit
overall factors of $\pm 1,i$, but we count different generating sets
(for the same code) and different orderings.\footnote{Our analysis revolves around random stabilizer generators rather
than random codes. As an aside, the resulting code is also randomly
distributed.  Also, any stabilizer of size $2^{n{-}k}$ has
$\prod_{b=0}^{n{-}k{-}1}\left(2^{n{-}k{-}b}{-}1\right)$ different
generating sets, so we have also found the total number of stabilizers
codes of this size.} There
are two constraints for the generating set, commutivity and
independence.
$S_1$ can be chosen from any of the $2^{2n}{-}1$ nontrivial Pauli
matrices.  Recall the property of $\cG_n$ stated in the previous
section.  $S_2$ can be chosen from the $2^{2n-1}$ Pauli matrices
commuting with $S_1$ but must be chosen from outside of the
multiplicative group generated by $S_1$, thus there are $2^{2n-1} - 2$
choices.  Similarly, each $S_i$ is chosen from the
$2^{2n{-}{(i{-}1)}}$ Pauli matrices commuting with $S_1, \cdots,
S_{i-1}$ but not from the multiplicative group generated by them, so
there are $2^{2n{-}(i{-}1)} - 2^{i{-}1}$ choices.  Thus, there are
$\Pi_{a=0}^{n{-}k{-}1} \; \left(2^{2n{-}a}{-}2^{a}\right)$ distinct
generating sets.

Now, for an arbitrary and fixed list $E_0, \cdots, E_{L}$ of
independent errors, how many choices of stabilizer generators will
give a code with $\{E_0^\dagger E_j\} \in N(S)$ $\forall
j{=}1,\cdots,L$?
This counting is similar to the above, but now $S_1,S_2\cdots S_{n-k}$ are
constrained to commute with $\{E_0^\dagger E_j\}$, in addition to the
two original constraints.
In other words, $S_1$ can be chosen from the $2^{2n-{L}}-1$ nontrivial
Pauli operators commuting with the $E_0^\dagger E_j$, and $S_2$ has
$2^{2n-{L+1}}-2$ choices, and so on.
Thus, there are $\Pi_{a=0}^{n{-}k{-}1} \;
\left(2^{2n-L-a}{-}2^{a}\right)$ sets of stabilizer generators that
commute with all of $E_0^\dg E_j$.

Putting together with the two stabilizer counts, one unconstrained and
the other with the same syndrome for $\{E_j\}_{j{=}0,\cdots,L}$, the
latter has probability
\begin{equation} 
  \frac
      {\Pi_{a=0}^{n{-}k{-}1} \; \left(2^{2n-L-a}{-}2^{a}\right)}
      {\Pi_{a=0}^{n{-}k{-}1} \; \left(2^{2n{-}a}{-}2^{a}\right)}
      \leq 2^{-L(n-k)}.
\end{equation} 

For step (2), we apply the union bound for the choice of the
$L{+}1$ $E_j$'s.  The probability that a random $[n,k]$ code is not
$L$-list decodable is less than $\binom{N_E}{L{+}1} {2^{-L(n{-}k)}}$,
which is no more than $N_E^{L{+}1} \,{2^{-L(n{-}k)}}$.  
The latter is less than $1$ if $k < n{-}(1{+}\smfrac{1}{L}) \log
N_E$.
But $N_E = |\cE^{pn}| = \sum_{r=0}^{\lfloor np \rfloor} 3^r {n \choose
r}$.  
For any $\delta>0$, $\exists n_\delta$ s.t. whenever $n\geq n_\delta$,
$\log N_E \leq n(H(p){+}p\log 3{+}\delta/3)$ so choosing
$k=n\left[1{-}\left(1{+}\smfrac{1}{L}\right)\left(H(p){+}p\log
3\right){-}2\delta/3\right]$ finishes the proof.
\end{proof} 

\bigskip

\section{Coding Strategy} 
Theorem \ref{Thm:ListCodesExist} tells us that for any $R <
1{-}H(p){-}p\log 3$, there exist $[n,Rn,pn,L]$-list codes for large
enough $n$ and $L$.  For example, we can choose the various parameters
as $\delta = 1 - H(p) - p\log 3 - R$, $L \geq
\smfrac{3}{\delta}(H(p){+}\log 3)$, and $n \geq n_\delta$ in
Thm.~\ref{Thm:ListCodesExist}).  Note that $L$ does not grow with $n$.

We now fix such a list-code, $\cC^{n,L}$.  This {\em always} returns a
syndrome $s$, a corresponding list of errors $Q^s_f \in \langle
P^s_l\rangle$, and a list-decoded state of the form $\sum_i
B_i^s\proj{\psi}{B_i^s}^\dg =:{\cal N}^s(\proj{\psi})$, where $\ket{\psi}$ is
the sender's intended logical state, $\sum_i {B^s_i}^\dg B^s_i = I$,
and each $B_i^s$ is in the span of $Q^s_f$.
Note that list-decoding removes all superposition between errors with
different syndromes.  Also, no approximation has been made so far.

Now we add a few more stabilizer generators to $\cC^{n,L}$ so that
with high probability (over the choice of the extra generators) the
receiver can decode $\ket{\psi}$ unambiguously.  These generators are
determined by a secret key shared by the sender and receiver, making
them unknown to the adversary.

It will follow from the proof of Thm.~\ref{Thm:PRCode} below that
adding $(1/\log(4/3))(2L+\log(1/\e))$ random generators to the code
$\cC^{n,L}$ would allow us to distinguish among the
$\{Q^s_f\}_{j=1}^{2^L}$ possible errors, with probability at least
$1{-}\e$.
This would require $2n(2L{+}\log(1/\e))/\log(4/3)$ bits of shared key.

A much smaller key can be used if small-biased sets are used to choose
these extra stabilizers pseudorandomly \cite{Naor90,Alon90}.
A subset of $\{0,1\}^{m}$, denoted $A$, is said to be an $\eta$-biased
set of {\em length m} if for each $e\in \{0,1\}^{m}$, roughly half of
the elements of $A$ have odd/even parity with $e$, or mathematically,
$\left|\Pr_{a\in A}\left( e\cdot a = 0\right) -\Pr_{a\in
A}\left(e\cdot a = 1\right) \right| \leq \eta$.
There are efficient constructions of $\eta$-biased sets of length $m$
with only $O(\frac{m^2}{\eta})$ elements \cite{Naor90,Alon90}.  

Let $G_0$ be the set of stabilizer generators of $\cC^{n,L}$.
We add $K$ extra stabilizer generators $T_{1},\cdots,T_{K}$.  When $j$
of these have been added, denote the code by $\cC^{n,L}_{j}$, with
$k{-}j$ encoded qubits and generator set $G_j$.  (Each
$\cC^{n,L}_{j}$ is a subcode of $\cC^{n,L}_{j{-}1}$.)
The next generator $T_{j{+}1}$ has to commute with all of 
$G_{j}$ but not be generated by it, thus, it is an encoded operation
on the code $\cC^{n,L}_{j}$.  Without loss of generality, it is an
encoded Pauli operation on the encoded $k{-}j$ qubits, and can be chosen
according to a random element of an $\eta$-biased set $A_{j{+}1}$ of
length $2(k{-}j)$.
The following theorem shows that using this procedure to add $K =
O(L\log 1/\e)$ stabilizers allows the receiver to reconstruct the
encoded state with high probability.
Using the efficient constructions of $\eta$-biased sets of length
$m\leq 2n$ with only $O(\frac{n^2}{\eta})$ elements, our construction
requires $O\left((2L+\log (1/\e))\log(n^2/\eta)\right)$ bits of key.

\begin{theorem}\label{Thm:PRCode}
Let $\cC^{n,L}$ be an $[n,Rn,pn,L]$-list code of rate $R$ and let
$\cC^{n,L}_{K}$ be the code obtained from $\cC^{n,L}$ by progressively
adding $K = (1/\log(4/3))(2L+\log(1/\e))$ stabilizers determined by
$\eta$-biased sets $A_{1},\cdots,A_{K}$ (of decreasing length) as
described above.  By using a secret key of fewer than $O(K(\log
(\frac{n^2}{\eta})))$ bits to select $\cC^{n,L}_{K}$, $nR- K =
n(R-o(n))$ qubits can be sent over $\N$ with fidelity at least $1-\e$
for all $\e < 1/2$.
\end{theorem}
\begin{proof}
The $[n,Rn,pn,L]$-list code reduces the adversary's power to choosing
 some ${\cal N}^s$ (with operation elements in the span of
 $\{Q^s_f\}_{f=1}^{2^L} = \langle P^s_l\rangle$) and a distribution of
 $s$.  
 So, if we prove that for each $s$, the probability (over the choice
 of $T_{1},\cdots,T_{K}$) is less than $\e$ to fail to distinguish
 between the $\{Q^s_f\}_{f=1}^{2^L}$, the fidelity of the decoded
 state with the original will be at least $1-\e$.
More specifically, fix an arbitrary $s$.  It is shown in \cite{NC96}
that $\cN^s$ has a $\chi$-representation $\cN^s(\rho) =
\sum_{f,f^\prime} \chi_{f, f^\prime} Q^s_f \rho (Q^s_{f^\prime})^\dg$
and let
\begin{equation}
F_s = \{k | \exists_{f,f^\prime}~\omega(T^k_l,Q_f^s) = \omega(T^k_l,Q_{f^\prime}^s) \}
\end{equation}
be the set of key values for which the additional stabilizer
generators fail to determine the list element.  Then, letting
$\proj{\psi_k}$ be the encoded logical state and $\cD^s_k$ be the
decoding operation given list $s$ and key $k$, our decoded state is
\begin{eqnarray}
& & \frac{1}{K} \sum_{k=1}^K \cD^s_k(\cN^s(\proj{\psi_k}))
\nonumber
\\
& = & \frac{1}{K}\sum_{k \not\in F_s}\cD^s_k(\cN^s(\proj{\psi_k})) +
\frac{1}{K}\sum_{k \in F_s}\cD^s_k(\cN^s(\proj{\psi_k}))
\nonumber
\\
&=& (1-\Pr(F_s))\proj{\psi} + \Pr(F_s)\phi_s,
\end{eqnarray}
where $\phi_s = \frac{1}{K \, \Pr(F_s)} \sum_{k \in F_s}
\cD^s_k(\cN^s(\proj{\psi_k}))$ is the state conditional on the key
failing to distinguish the list elements properly.  We will now show
that $\Pr(F_s)$ can be made less than $\e$ for all lists of length
$2^L$ by choosing $K$ as in the theorem.  This results in a decoding
fidelity of at least $1 - \e$.

Now fix $f,f'$ and define the events $M_j$ as $\{\omega(Q^s_{f}, T_j)
= \omega(Q^s_{f'}, T_j)\}$.
Then, the probability, over the choices of $T_{1,\cdots,K}$, that they
assign the same syndrome to $Q^s_{f}$ and $Q^s_{f'}$ is $\Pr\left(
\cap_{j=1}^{K} M_j\right) = \prod_{j=1}^{K}\Pr\left(M_j|M_{j-1}\dots
M_1\right)$.
Since each $T_j$ is chosen using an $\eta$-biased string of encoded
operations of the code $\cC^{n,L}_{j{-}1}$, we have
$\Pr\left(M_j|M_{j-1}\dots M_1\right) \leq \smfrac{1+\eta}{2}$, which
immediately implies that $\Pr\left( \cap_{j=1}^{K} M_j\right) \leq
\left(\frac{1+\eta}{2}\right)^K$.  
By a union bound over the choice of $f,f'$, the probability of {\em
any} pair having the same commutation relations for all $j$ is less
than $2^{2L}\left(\frac{1+\eta}{2}\right)^K$.

By choosing $\eta \leq 1/2$, $K = (1/\log(4/3))(2L+\log(1/\e))$ we
make this failure probability less than $\e$ so that with probability
at least $1-\e$, $Q^s_f$ can be unambiguously identified and the state
reconstructed.  
\end{proof}
Note that $\e$ can be made to vanish exponentially with $n$ without
incurring extra $n$-dependence on the key size.  

In Theorem \ref{Thm:PRCode}, the extra generators can distinguish the
worst case ${\cal N}^s$ and no union bound over $s$ is needed.
%
Furthermore, the additional stabilizers do {\em not} depend on $s$.
It means that the final construction is a single quantum error correcting
code depending only on a small key.  It also means that it is not
necessary to first perform list-decoding before selecting the extra
stabilizers.  The combined decoding operation is independent of $s$
but concludes the error based on the joint inputs of $s$ and the extra
syndrome bits (one possible way of which is to first output a list
based on $s$).

\bigskip

\section{Discussion} 
\label{sec:disc}
We have introduced the adversarial quantum channel and shown that
using a logarithmic length secret key one can communicate over this
channel with a rate of $1{-}H(p){-}p\log 3$.  This is far higher than
would be naively expected from existing QECCs, and
quite close to the best known rates for
%
independent depolarizing channels of error probability $p$.
Our construction involves quantum list codes, which we defined
and showed to exist with favorable parameters.  Classical list decoding
has recently played an important role in several complexity theoretic results (for a review, see \cite{Sudan00}),
and we expect quantum list codes will be similarly useful in the context of quantum complexity theory.  

The scenario considered in this paper and the spirit of our protocols
are closely related to those of \cite{CGS05}.  Comparing their result
with ours points to interesting open questions.  
Reference \cite{CGS05} constructed {\em approximate} quantum error
correcting codes of length $n$ capable of correcting up to $(n-1)/2$
errors with high probability (compared to at most $n/4$ correctable 
errors for an exact code).  
Thus, the fraction of errors that can be tolerated in \cite{CGS05}
approaches $1/2$ as $n$ gets large, which is much higher than in our
scheme.  Furthermore, unlike our scheme, no secret key is
required.  Instead, randomizing parameters are sent as part of the
message via carefully constructed secret sharing schemes.  
However, the alphabet size of the codes in \cite{CGS05} grows as a
function of both the blocklength and the code's accuracy which
severely limits the transmission rate.  Also, when their large
dimensional channel is viewed as a block of qubit channels, the
adversary considered in \cite{CGS05} is much more restricted than
ours, being limited to the corruption of {\em contiguous blocks} of
qubits.

Altogether, there is a general open question on the tradeoff between
distance, rate, and key required for a code.
More specifically, 
it is an interesting question whether there are {\em qubit}
approximate QECCs which achieve the rates of our codes without using a
secret key, or, less ambitiously, one with constant size.
We have also left unanswered the capacity of $\N$ assisted by a
negligible length secret key.  
It seems plausible that the capacity is equal to that of the
depolarizing channel with error rate $p$, which would be in analogy
with the classical result of \cite{Lang04}.  
While the capacity of for the depolarizing channel is an open
question, one may find codes for $\N$ with rates matching the best
known for the depolarizing channel.
It will also be interesting to consider other side resources such as a
negligible amount of entanglement.  Finally, unlike DMCs, it is
unclear for adversarial channels whether the capacity can be improved
with a small number of uses of noiseless quantum channels.

As a side remark, our scheme uses the secret key as a
randomizing parameter that is inaccessible to the adversary.  Since
the adversary must corrupt the transmitted state before it is received
by Bob, if Bob is allowed to send a ``receipt'' of the quantum states
to Alice, she can simply disclose the random code afterwards and no
key is required.  In other words, one bit of back communication
along with logarithmic forward classical communication (all
authenticated) can replace the key requirement.

As another side remark, as the channel can be used to create
entanglement, the key used in the communication can be replenished 
by sending a negligible number of EPR pairs (without affecting the 
communication rate).  Thus the key requirement is only catalytic.  

Our result also finds application to a related problem---entanglement
distillation with bounded weight errors.  In this problem, a state is
already distributed between Alice and Bob, so the adversary has
already acted and randomizing parameters can be sent in public without
a receipt.  In \cite{AG03}, it was shown that $n$ noisy EPR pairs
with errors of weight up to $pn$ could be purified to $n(1-H(p)-p\log
3)$ perfect EPR pairs by a two-way distillation procedure.  Our
construction lets us distill high fidelity EPR pairs at the same rate
with only forward classical communication.  In fact, it was suggested
in \cite{AG03} that quantum list codes could be used to reduce the
computational complexity of their protocols---almost exactly the
approach taken here, though in our case with an eye towards reducing
the communication required.  The question of efficent encoding and
decoding via list codes has not yet been resolved.

It may also be interesting to consider how restricting the
computational power of our adversary affects the channel's capacity,
which is another topic we leave to future work.  The investigation of
other restrictions (such as causality of the adversarial channel) is
also natural in certain situations and may lead to additional
insights.

\smallskip

\section*{Acknowledgments} 
We are grateful to Roberto Oliveira, John Smolin, Daniel Gottesman,
and especially Aram Harrow for helpful discussions, as well as to the
IBM T.J. Watson Research Center, where the bulk of this work was
completed.  DL acknowledges travel funds from the CIAR to visit IBM,
and funding from the Tolman Foundation, CIAR, NSERC, CRC, CFI, and
OIT. GS received support from the US NSF grant PHY-0456720 and
Canada's NSERC.  We both received support from US NSF grant
EIA-0086038.



\bibliographystyle{IEEEbib}

\begin{thebibliography}{10}

\bibitem{D03}
I.~Devetak,
\newblock ``The private classical capacity and quantum capacity of a quantum
  channel,''
\newblock {\em IEEE Trans. Inf. Theory}, vol. 51, pp. 44--55, 2005,
\newblock ar{X}iv:quant-ph/0304127.

\bibitem{DHW03}
I.~Devetak, A.~W. Harrow, and A.~Winter,
\newblock ``A family of quantum protocols,''
\newblock {\em Phys. Rev. Lett.}, vol. 93, pp. 230505, 2004,
\newblock ar{X}iv quant-ph/0308044.

\bibitem{DS03}
I.~Devetak and P.~Shor,
\newblock ``The capacity of a quantum channel for simultaneous transmission of
  classical and quantum information,''
\newblock {\em Comm. Math. Phys.}, vol. 256, no. 2, pp. 287--303, 2005,
\newblock ar{X}iv quant-ph/0311131.

\bibitem{Shor02}
P.W. Shor,
\newblock ``The quantum channel capacity and coherent information.,''
\newblock lecture notes, MSRI Workshop on Quantum Computation, 2002.
  http://www.msri.org/publications/ \\ ln/msri/2002/quantumcrypto/shor/1/.

\bibitem{W99}
A.~Winter,
\newblock ``Coding theorem and strong converse for quantum channels,''
\newblock {\em IEEE Trans. Inf. Theory}, vol. 45, pp. 2481--2485, 1999.

\bibitem{W04POVM}
A.~Winter,
\newblock ````\mbox{E}xtrinsic'' and ``intrinsic'' data in quantum
  measurements: asymptotic convex decomposition of positive operator valued
  measures,''
\newblock {\em Comm. Math. Phys.}, vol. 244, pp. 157--185, 2004.

\bibitem{BarnumKN00}
H.~Barnum, E.~Knill, and M.~A. Nielsen,
\newblock ``On quantum fidelities and channel capacities,''
\newblock {\em IEEE Trans. Inf. Theory}, vol. 46, pp. 1317--1329, 2000.

\bibitem{SW97}
B.~Schumacher and M.~Westmoreland,
\newblock ``Sending classical information via noisy quantum channels,''
\newblock {\em Phys. Rev. A}, vol. 56, pp. 131--138, 1997.

\bibitem{Holevo98}
A.~Holevo,
\newblock ``The capacity of quantum channel with general signal states,''
\newblock {\em IEEE Trans.Info.Theor.}, vol. 44, pp. 269--273, 1998,
\newblock ar{X}iv:quant-ph/9611023.

\bibitem{Lloyd97}
S.~Lloyd,
\newblock ``Capacity of the noisy quantum channel,''
\newblock {\em Phys. Rev. A}, vol. 55, pp. 1613--1622, 1997.

\bibitem{BSST02}
C.H. Bennett, P.W. Shor, J.A. Smolin, and A.V.Thapliyal,
\newblock ``Entanglement-assisted capacity of a quantum channel and the reverse
  shannon theorem,''
\newblock {\em IEEE Trans. Inf. Theory}, vol. 48, pp. 2637--2655, 2002.

\bibitem{BDM03}
Garry Bowen, Igor Devetak, and Stefano Mancini,
\newblock ``Bounds on classical information capacities for a class of quantum
  memory channels,''
\newblock {\em Phys.Rev. A.}, vol. 71, pp. 034310, 2005,
\newblock ar{X}iv quant-ph/0312216.

\bibitem{KW05}
Dennis Kretschmann and Reinhard~F. Werner,
\newblock ``Quantum channels with memory,''
\newblock {\em Phys. Rev. A}, vol. 72, pp. 062323, 2005,
\newblock ar{X}iv:quant-ph/0502106.

\bibitem{HN02}
Masahito Hayashi and Hiroshi Nagaoka,
\newblock ``General formulas for capacity of classical-quantum channels,''
\newblock {\em IEEE Trans. Inf. Theory}, vol. 49, pp. 1753--1768, 2003,
\newblock ar{X}iv quant-ph/0206186.

\bibitem{RenThesis}
Renato Renner,
\newblock ``Security of quantum key distribution,''
\newblock Ph.d. Thesis, Swiss Federal Institute of Technology, 2005.

\bibitem{BD06}
G.~Bowen and N.~Datta,
\newblock ``Beyond i.i.d. in quantum information theory,''
\newblock ar{X}iv quant-ph/0604013.

\bibitem{Lang04}
M.~Langberg,
\newblock ``Private codes or succinct random codes that are (almost) perfect,''
\newblock {\em Proceedings of FOCS}, pp. 325--334, 2004.

\bibitem{Guru03}
V.~Guruswami,
\newblock ``List decoding with side information,''
\newblock {\em IEEE Conference on Computational Complexity}, pp. 300--309,
  2003.

\bibitem{Hamming50}
R.W. Hamming,
\newblock ``Error detecting and error correcting codes,''
\newblock {\em Bell System Technical Journal}, vol. 29, pp. 147--160, 1950.

\bibitem{Gottesman97}
D.~Gottesman,
\newblock ``Stabilizer codes and quantum error correction,''
\newblock {C}altech Ph.D. Thesis.

\bibitem{Rains99}
E.~Rains,
\newblock ``Quantum shadow enumerators,''
\newblock {\em IEEE Trans.Info.Theor.}, vol. 45, pp. 2361--2366, 1999,
\newblock ar{X}iv: quant-ph/9611001.

\bibitem{Rains03}
E.~Rains,
\newblock ``New asymptotic bounds for self-dual codes and lattices,''
\newblock {\em IEEE Trans.Info.Theor.}, vol. 49, pp. 1261--1274, 2003.

\bibitem{BDSW96}
C.~H. Bennett, D.~P. DiVincenzo, J.~A. Smolin, and W.~K. Wootters,
\newblock ``Mixed state entanglement and quantum error correction,''
\newblock {\em Phys.Rev. A.}, vol. 54, pp. 3824--3851, 1996,
\newblock ar{X}iv quant-ph/9604024.

\bibitem{ShorPreskill00}
P.W. Shor and J.~Preskill,
\newblock ``Simple proof of security of the bb84 quantum key distribution
  protocol,''
\newblock {\em Phys. Rev. Lett.}, vol. 85, pp. 441--444, 2000.

\bibitem{Leung97b}
D.~Leung, M.~Nielsen, I.~Chuang, and Y.~Yamamoto,
\newblock ``Approximate quantum error correction can lead to better codes,''
\newblock {\em Phys. Rev. A}, vol. 56, pp. 2567--2573, 1997,
\newblock \mbox{arXive} e-print quant-ph/9704002.

\bibitem{Schumacher01}
B.~Schumacher and M.~D. Westmoreland,
\newblock ``Approximate quantum error correction,''
\newblock {\em Quantum information Processing}, vol. 1, 2002,
\newblock quant-ph/0112106.

\bibitem{CGS05}
C.~Crepeau, D.~Gottesman, and Adam Smith,
\newblock ``Approximate quantum error-correcting codes and secret sharing
  schemes,''
\newblock {\em Advances in Cryptology -- EUROCRYPT}, 2005,
\newblock ar{X}iv quant-ph/0503139.

\bibitem{QIP06}
Graeme Smith,
\newblock ``Communicating over adversarial quantum channels,''
\newblock In QIP 2006, Paris. http://www.lri.fr/qip06/slides/ smith.pdf, 2006.

\bibitem{KY06}
Akinori Kawachi and Tomoyuki Yamakami,
\newblock ``Quantum hardcore functions by complexity-theoretical quantum list
  decoding,''
\newblock ar{X}iv:quant-ph/0602088.

\bibitem{Hayashi06}
Masahito Hayashi,
\newblock ``Channel capacities of classical and quantum list decoding,''
\newblock ar{X}iv:quant-ph/0603031.

\bibitem{Nielsen00bk}
Michael Nielsen and Issac Chuang,
\newblock {\em Quantum computation and quantum information},
\newblock Cambridge, 2004.

\bibitem{Naor90}
Joseph Naor and Moni Naor,
\newblock ``Small-bias probability spaces: Efficient constructions and
  applications,''
\newblock in {\em {ACM} Symposium on Theory of Computing}, 1990, pp. 213--223.

\bibitem{Alon90}
Noga Alon, Oded Goldreich, Johan Hastad, and Rene Peralta,
\newblock ``Simple constructions of almost k-wise independent random
  variables,''
\newblock in {\em {IEEE} Symposium on Foundations of Computer Science}, 1990,
  pp. 544--553.

\bibitem{NC96}
M.~Nielsen and I.~Chuang,
\newblock ``Prescription for experimental determination of the dynamics of a
  quantum black box,''
\newblock ar{X}iv:quant-ph/9610001.

\bibitem{Sudan00}
Madhu Sudan,
\newblock ``List decoding: Algorithms and applications,''
\newblock {\em Lecture Notes in Computer Science}, vol. 1872, pp. 25, 2000.

\bibitem{AG03}
A.~Ambainis and D.~Gottesman,
\newblock ``The minimum distance problem for two-way entanglement
  purification,''
\newblock {\em IEEE Trans. Inf. Theory}, vol. 52, pp. 748--753, 2006,
\newblock ar{X}iv quant-ph/0310097.

\end{thebibliography}

\raggedbottom 



\end{document}